\documentclass[conference]{IEEEtran}
\IEEEoverridecommandlockouts

\usepackage{cite}
\usepackage{amsmath,amssymb,amsfonts}
\usepackage{amsthm}
\usepackage{algorithmic}
\usepackage{graphicx}
\usepackage{textcomp}
\usepackage{xcolor}

\theoremstyle{plain}
\newtheorem{theorem}{Theorem}
\newtheorem{lemma}{Lemma}
\newtheorem{proposition}{Proposition}
\newtheorem{corollary}{Corollary}

\theoremstyle{definition}
\newtheorem{definition}{Definition}
\newtheorem{example}{Example}
\newtheorem{remark}{Remark}

\begin{document}
\bstctlcite{IEEEexample:BSTcontrol}
\title{
    On Diagonalizable
    Delay--Doppler Channels and Their Diagonalizing Waveforms
}

\author{
\IEEEauthorblockN{Sirui Li}
\IEEEauthorblockA{
Fudan University\\
Shanghai, China\\
Email: 25213090137@m.fudan.edu.cn
}
\and
\IEEEauthorblockN{Cheng Du}
\IEEEauthorblockA{Great Bay University\\
Dongguan, China\\
Email: cdu@gbu.edu.cn
}
\and
\IEEEauthorblockN{Yu Zhu}
\IEEEauthorblockA{
Fudan University\\
Shanghai, China\\
Email: zhuyu@fudan.edu.cn
}
}

\maketitle
\begin{abstract}
In doubly selective channels, the joint delay and Doppler dispersion generally induces coupling among transmitted symbols, thereby increasing receiver equalization complexity.
Nevertheless, by using
appropriately designed waveforms, channels with
certain delay-Doppler (DD) supports can be diagonalized for one-tap equalization. The whole picture of such DD supports and their corresponding waveforms is still unclear, except for several examples identified in literature. In this paper, under cyclic-prefix (CP)-based block transmission and assuming that the modulation waveforms form an orthonormal basis, we identify all such channel supports by an elementary expression, and derive the corresponding
waveforms in closed form.
\end{abstract}

\begin{IEEEkeywords}
Delay--Doppler channels, waveform design, channel diagonalization.
\end{IEEEkeywords}

\section{Introduction}
Waveform design is a fundamental issue in the development of new
wireless systems. Future networks are expected to support high
mobility, wide bandwidths, and integrated sensing and communication
(ISAC), placing increasingly stringent demands on waveform robustness
to delay-Doppler (DD) dispersion
\cite{tataria2021sixg,liu2022isac}.

Existing waveforms for doubly selective channels can be broadly classified into two categories. The first class aims to exploit the channel diversity, including the 
orthogonal time frequency space (OTFS) and affine frequency division multiplexing (AFDM). OTFS \cite{hadani2017otfs} and its related DD-domain
waveforms, including orthogonal delay-doppler division multiplexing (ODDM) and Zak-OTFS \cite{lin2022oddm,mattu2026zakotfs}, place information symbols in the DD domain and map them to the time--frequency domain. Note that vector orthogonal frequency division
multiplexing (V-OFDM) shares the same waveform as OTFS, though developed from a different perspective \cite{xia2001vofdm,xia2022otfsvofdm}. AFDM uses the discrete affine Fourier transform (DAFT) and selects its parameters to separate different paths in the effective channel \cite{bemani2022afdm}. Although OTFS and AFDM improve performance by exploiting channel
diversity \cite{surabhi2019diversity,bemani2022afdm,xia2025rethink},
they incur higher equalization and detection complexity
\cite{raviteja2018otfs,bemani2022equalization}.

The second class, represented by DAFT-OFDM, pursues
low-complexity equalization through channel diagonalization. OFDM
enables one-tap equalization in delay-only multipath channels
\cite{wang2000multicarrier,matz2013tff}, whereas DAFT-OFDM eliminates
intercarrier interference when the path delays and Doppler shifts
satisfy a linear relation \cite{erseghe2005aft}. For more general
channels, approximate diagonalization provides an alternative
\cite{kozek1998}. Although these waveforms may exploit
less channel diversity than the first class in uncoded systems, diversity can instead be
provided through coding and interleaving, as in coded OFDM, while
retaining simple equalization
\cite{caire1998bit_interleaved,akay2006full_diversity}.

In this work, we adopt the design philosophy of the second class and focus waveform design on channel diagonalization and low-complexity equalization. With full channel state information, singular value decomposition (SVD)-based precoding and combining can always create parallel subchannels \cite[Ch.~7]{tse2005fundamentals}. In waveform design, however, the individual path gains are generally unknown, and no fixed waveform can diagonalize all possible channels over the \emph{entire DD plane}, as a consequence of the fundamental
delay--Doppler uncertainty \cite{matz2013tff}. Nevertheless, exact diagonalization remains possible for \emph{certain subsets of DD supports} without knowledge of the path gains, as illustrated by OFDM, DAFT-OFDM, and Zak-OTFS
\cite{wang2000multicarrier,matz2013tff,erseghe2005aft,mattu2026zakotfs}.

This motivates a fundamental question: which DD supports admit such diagonalization, and what are the corresponding waveforms? We address this question under cyclic-prefix (CP)-based block transmission with a
square modulation matrix whose columns form an orthonormal basis.

Our contributions are twofold. First, we give a simple, explicit
representation of all maximal diagonalizable DD supports. This
representation shows that rectangular grids and DD lines are special
cases of a broader family that also includes sheared grids, and it
also gives the exact number of such supports. Second, for each
identified support, we derive in closed form an orthonormal waveform
basis that diagonalizes all channels with that support. We further
show that the resulting modulation can be implemented using parallel
FFTs and phase rotations. Zak-OTFS and DAFT-OFDM are
obtained as special cases of this construction.

The remainder of this paper is organized as follows.
Section~II presents the DD channel model and the
diagonalization criterion. Section~III identifies all maximal diagonalizable DD supports. Section~IV derives their corresponding waveforms. Section~V concludes the paper.

\emph{Notation:}
Lowercase italic letters denote scalars, bold lowercase letters denote
column vectors, and bold uppercase letters denote matrices. Unless
stated otherwise, indices start at zero. The superscripts
\((\cdot)^T\) and \((\cdot)^H\) denote transpose and Hermitian
transpose, respectively. The symbols \(\mathbf{I}_K\) and
\(\mathbf{0}\) denote the \(K\times K\) identity matrix and an
all-zero vector or matrix, respectively, with the dimension of
\(\mathbf{0}\) indicated by a subscript or inferred from context.
The sets of integers and complex numbers are denoted by
\(\mathbb{Z}\) and \(\mathbb{C}\), respectively, and
\(\mathrm{i}=\sqrt{-1}\). For a positive integer \(K\),
\(\mathbb{Z}_K\triangleq\mathbb{Z}/K\mathbb{Z}\) denotes the ring
of integers modulo \(K\), identified with
\(\{0,\ldots,K-1\}\). Arithmetic in \(\mathbb{Z}_K\) is performed
modulo \(K\), and \([a]_K\in\mathbb{Z}_K\) denotes \(a\) reduced
modulo \(K\). The notation \(a\equiv b\pmod K\) denotes congruence
modulo \(K\), and \(d\mid K\) means that \(d\) divides \(K\). For
\(\boldsymbol{g}_1,\ldots,\boldsymbol{g}_r\in\mathbb{Z}_K^2\),
\(\langle\boldsymbol{g}_1,\ldots,\boldsymbol{g}_r
\rangle_{\mathbb{Z}_K}
\triangleq
\{\sum_{q=1}^{r}a_q\boldsymbol{g}_q:
a_q\in\mathbb{Z}_K\}\)
denotes their generated \(\mathbb{Z}_K\)-submodule. For a scalar \(z\)
and a finite set \(\mathcal{S}\), \(|z|\) and \(|\mathcal{S}|\)
denote magnitude and cardinality, respectively. The symbols
\(\otimes\) and \(\operatorname{vec}(\cdot)\) denote the Kronecker
product and column-wise vectorization, respectively.

\section{System Model and Diagonalization Criterion}
\label{sec:system_model}
\subsection{Unified Transceiver and DD Channel Model}
\label{subsec:unified_transceiver_model}
Although different waveforms may use different guard structures, we
focus on cyclic-prefix (CP)-based block transmission. The CP length is
assumed to be no shorter than the maximum channel delay, so that
interblock interference can be neglected after CP removal. At the transmitter, let
\(\mathbf{s}=[s_0,\ldots,s_{N-1}]^T\in\mathcal{A}^N\), where \(\mathcal{A}\subseteq\mathbb{C}\) denotes the symbol alphabet. Let
\(\mathbf{B}=[\boldsymbol{\phi}_0,\ldots,
\boldsymbol{\phi}_{N-1}]\in\mathbb{C}^{N\times N}\) satisfy
\(\mathbf{B}^H\mathbf{B}=\mathbf{I}_N\). The transmitted block is
\begin{equation}
    \mathbf{x}
    =
    \mathbf{B}\mathbf{s}
    =
    \sum_{m=0}^{N-1}s_m\boldsymbol{\phi}_m.
    \label{eq:modulation_model}
\end{equation}

The interpretation of the entries of \(\mathbf{s}\) depends on the
chosen waveform. For Zak-OTFS, \(\mathbf{s}\) contains DD-domain symbols, whereas for DAFT-OFDM it contains
DAFT-domain symbols. In either case, the modulation matrix
\(\mathbf{B}\) maps these symbols to the time-domain block
\(\mathbf{x}\)
\cite{mattu2026zakotfs,erseghe2005aft}.
A CP is then appended to \(\mathbf{x}\) before transmission.

We next consider a \(P\)-path doubly selective channel. A standard
specular multipath model represents its continuous DD spreading
function as
\cite{bello1963,matz2013tff}
\begin{equation}
    S_H(\tau,\nu)
    =
    \sum_{p=0}^{P-1}
    h_p
    \delta(\tau-\tau_p)
    \delta(\nu-\nu_p),
    \label{eq:continuous_dd_spreading_function}
\end{equation}
where \(\delta(\cdot)\) is the Dirac delta function, and
\(h_p\), \(\tau_p\), and \(\nu_p\) denote the complex gain, delay,
and Doppler shift of the \(p\)th path, respectively.

After receive filtering, sampling, and CP removal, the continuous
channel in \eqref{eq:continuous_dd_spreading_function} induces an
equivalent channel matrix
\(\mathbf{H}\in\mathbb{C}^{N\times N}\) over one \(N\)-sample block.
Let \(\zeta_N\triangleq e^{2\pi\mathrm{i}/N}\). For a discrete DD
index \(\boldsymbol{u}=(\tau,\nu)\in\mathbb{Z}_N^2\), let
\(\mathbf{T}_{\tau}\) and \(\mathbf{M}_{\nu}\) denote the cyclic
time-shift and Doppler-shift matrices, respectively, whose entries are
\begin{equation}
    [\mathbf{T}_{\tau}]_{n,m}
    =
    \delta_{m,[n-\tau]_N},
    \qquad
    [\mathbf{M}_{\nu}]_{n,m}
    =
    \zeta_N^{\nu n}\delta_{m,n},
    \quad
    n,m\in\mathbb{Z}_N.
    \label{eq:delay_doppler_shift_matrices}
\end{equation}
Here, \(\delta_{a,b}\) denotes the Kronecker delta.
The corresponding Weyl operator is
\cite{kozek1998,matz2013tff}
\begin{equation}
    \mathbf{W}_N(\boldsymbol{u})
    \equiv
    \mathbf{W}_N(\tau,\nu)
    \triangleq
    \mathbf{T}_{\tau}\mathbf{M}_{\nu},
    \label{eq:weyl_operator_matrix}
\end{equation}
and therefore
\begin{equation}
    \bigl[
        \mathbf{W}_N(\tau,\nu)\mathbf{x}
    \bigr][n]
    =
    \zeta_N^{\nu(n-\tau)}
    x\bigl([n-\tau]_N\bigr),
    \qquad
    n\in\mathbb{Z}_N.
    \label{eq:weyl_operator}
\end{equation}
Using Weyl-operator expansion
\cite{kozek1998,matz2013tff}, the equivalent sampled channel is
expressed in the discrete DD domain as
\begin{equation}
    \mathbf{H}
    =
    \sum_{\boldsymbol{u}\in\mathcal{S}}
    h[\boldsymbol{u}]
    \mathbf{W}_N(\boldsymbol{u}),
    \label{eq:channel_weyl_expansion}
\end{equation}
where
\(\mathcal{S}
\triangleq
\{\boldsymbol{u}\in\mathbb{Z}_N^2:
h[\boldsymbol{u}]\neq0\}\)
is the discrete DD support and \(h[\boldsymbol{u}]\) is the associated
spreading coefficient.

Passing the transmit block in \eqref{eq:modulation_model} through the
channel in \eqref{eq:channel_weyl_expansion} and adding white Gaussian
noise yields
\begin{equation}
    \mathbf{y}
    =
    \mathbf{H}\mathbf{x}
    +
    \mathbf{w}
    =
    \mathbf{H}\mathbf{B}\mathbf{s}
    +
    \mathbf{w},
    \qquad
    \mathbf{w}
    \sim
    \mathcal{CN}
    \left(
        \mathbf{0},
        \sigma_w^2\mathbf{I}_N
    \right).
    \label{eq:received_signal}
\end{equation}

Here, \(\mathcal{CN}(\boldsymbol{\mu},\mathbf{C})\) denotes a proper
complex Gaussian distribution with mean \(\boldsymbol{\mu}\) and
covariance matrix \(\mathbf{C}\).

The receiver then applies the matched-filter bank
\(\mathbf{B}^{H}\), giving
\begin{equation}
    \mathbf{z}
    =
    \mathbf{B}^{H}\mathbf{y}
    =
    \mathbf{B}^{H}\mathbf{H}\mathbf{B}\mathbf{s}
    +
    \widetilde{\mathbf{w}},
    \qquad
    \widetilde{\mathbf{w}}
    \triangleq
    \mathbf{B}^{H}\mathbf{w}.
    \label{eq:matched_filter_output}
\end{equation}
Since \(\mathbf{B}\) is unitary, the filtered noise remains white, with
\(\widetilde{\mathbf{w}}
\sim
\mathcal{CN}(\mathbf{0},\sigma_w^2\mathbf{I}_N)\).
Hence, \(\mathbf{B}^{H}\mathbf{H}\mathbf{B}\) is the effective channel
matrix seen by the modulation symbols.
\subsection{Diagonalizable DD Supports}
When the effective channel matrix
$\mathbf{B}^{H}\mathbf{H}\mathbf{B}$ is diagonal, the modulation
symbols are decoupled and can be independently recovered using
one-tap equalization. This motivates the following definition.

\begin{definition}
\label{def:diagonalizable_dd_support}
A DD support
$\mathcal{S}\subseteq\mathbb{Z}_N^2$
is \emph{diagonalizable} if there exists a unitary matrix $\mathbf{B}$, depending only on $\mathcal{S}$, such that
$\mathbf{B}^{H}\mathbf{H}\mathbf{B}$ is diagonal for every channel whose DD support is contained in \(\mathcal S\).
\end{definition}
Substituting \eqref{eq:channel_weyl_expansion} into
$\mathbf{B}^{H}\mathbf{H}\mathbf{B}$ yields
$\mathbf{B}^{H}\mathbf{H}\mathbf{B}
=
\sum_{\boldsymbol{u}\in\mathcal{S}}
h[\boldsymbol{u}]
\mathbf{B}^{H}\mathbf{W}_N(\boldsymbol{u})\mathbf{B}$.
Since the coefficients $h[\boldsymbol{u}]$ can take arbitrary
values, the diagonalizability condition reduces to requiring
$\mathbf{B}^{H}\mathbf{W}_N(\boldsymbol{u})\mathbf{B}$ to be
diagonal for every $\boldsymbol{u}\in\mathcal{S}$. Equivalently, the
columns of $\mathbf{B}$ form a common orthonormal eigenbasis of the
Weyl operators indexed by $\mathcal{S}$.
The following lemma characterizes when such a unitary matrix
$\mathbf{B}$ exists 
\cite[Th.~2.5.5]{horn2012matrix}.

\begin{lemma}[Common eigenbasis]
\label{lem:common_orthonormal_eigenbasis}
Let $\mathcal{S}\subseteq\mathbb{Z}_N^2$ be nonempty. The Weyl
operators
$\{\mathbf{W}_N(\boldsymbol{u}):
\boldsymbol{u}\in\mathcal{S}\}$
admit a common orthonormal eigenbasis if and only if they commute
pairwise, that is,
\begin{equation}
    \mathbf{W}_N(\boldsymbol{u})
    \mathbf{W}_N(\boldsymbol{v})
    =
    \mathbf{W}_N(\boldsymbol{v})
    \mathbf{W}_N(\boldsymbol{u}),
    \qquad
    \forall\,
    \boldsymbol{u},\boldsymbol{v}\in\mathcal{S}.
    \label{eq:common_eigenbasis_condition}
\end{equation}
\end{lemma}

\begin{figure*}[!t]
    \centering
    \includegraphics[
        width=\textwidth,
        keepaspectratio
    ]{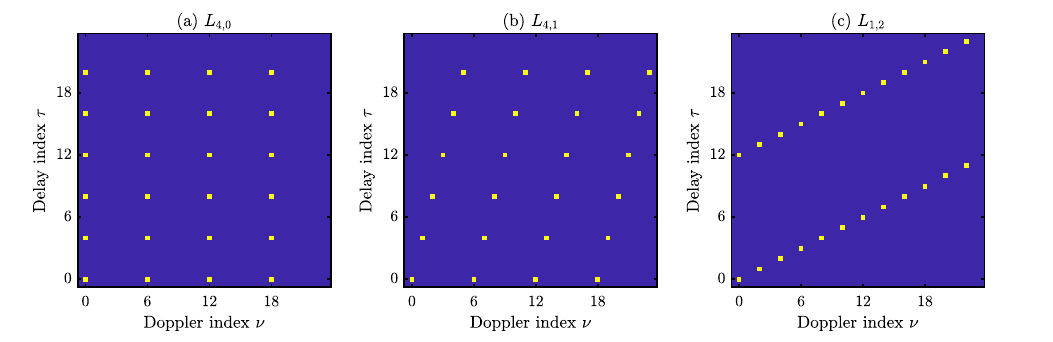}
    \caption{
    Representative Lagrangian submodules in the DD plane for
    $N=24$: (a) the rectangular grid $L_{4,0}$,
    (b) the sheared grid $L_{4,1}$, and
    (c) the cyclic line $L_{1,2}$.
    }
    \label{fig:lagrangian_submodule_geometry}
\end{figure*}

For Weyl operators, the commutativity can be expressed
directly in terms of the DD indices. For
$\boldsymbol{u}=(\tau,\nu)$ and
$\boldsymbol{v}=(\tau',\nu')$, define
\begin{equation}
    \omega(\boldsymbol{u},\boldsymbol{v})
    \triangleq
    \bigl[
        \tau\nu'
        -
        \nu\tau'
    \bigr]_N.
    \label{eq:dd_commutation_measure}
\end{equation}
The corresponding Weyl operators satisfy
\begin{equation}
    \mathbf{W}_N(\boldsymbol{u})
    \mathbf{W}_N(\boldsymbol{v})
    =
    \zeta_N^{-\omega(\boldsymbol{u},\boldsymbol{v})}
    \mathbf{W}_N(\boldsymbol{v})
    \mathbf{W}_N(\boldsymbol{u}).
    \label{eq:weyl_commutation}
\end{equation}
Hence, they commute if and only if
\begin{equation}
    \omega(\boldsymbol{u},\boldsymbol{v})=0.
    \label{eq:weyl_commuting_condition}
\end{equation}

Combining this relation with
Lemma~\ref{lem:common_orthonormal_eigenbasis}
gives the following criterion.

\begin{proposition}[Diagonalization criterion]
\label{prop:diagonalization_criterion}
A nonempty DD support
$\mathcal{S}\subseteq\mathbb{Z}_N^2$
is diagonalizable if and only if
\begin{equation}
    \omega(\boldsymbol{u},\boldsymbol{v})=0,
    \qquad
    \forall\,
    \boldsymbol{u},\boldsymbol{v}\in\mathcal{S}.
    \label{eq:dd_support_diagonalization_condition}
\end{equation}
\end{proposition}

Any DD set satisfying the pairwise commutation condition contains at
most $N$ elements \cite{albouy2009isotropic}. Definition~\ref{def:lagrangian_dd_set} describes the case in which this upper bound is
attained.

\begin{definition}[Lagrangian submodule]
\label{def:lagrangian_dd_set}
Following \cite{albouy2009isotropic}, a submodule
$L\subseteq\mathbb{Z}_N^2$ is called a
\emph{Lagrangian submodule} if
\begin{equation}
    \omega(\boldsymbol{u},\boldsymbol{v})=0,
    \quad
    \forall\,\boldsymbol{u},\boldsymbol{v}\in L,
    \qquad
    |L|=N.
\end{equation}
\end{definition}

By Definition~\ref{def:lagrangian_dd_set}, Lagrangian submodules are
exactly the maximal diagonalizable DD supports. We next derive an elementary representation of Lagrangian submodules without invoking the Chinese remainder
theorem used in \cite{albouy2009isotropic}.

\section{Identification of Maximal Diagonalizable DD Supports}
\label{sec:classification}
\subsection{Elementary Representation}

\begin{theorem}[Maximal diagonalizable DD supports]
\label{thm:elementary_representation}
Let
\begin{equation}
    \mathcal{L}_N
    \triangleq
    \left\{
        L\subseteq\mathbb{Z}_N^2:
        L \text{ is a Lagrangian submodule}
    \right\}
\end{equation}
be the family of all Lagrangian submodules of
$\mathbb{Z}_N^2$, and define
\begin{equation}
    \mathcal{E}_N
    \triangleq
    \left\{
        L_{d,\beta}:
        d\mid N,\;
        \beta\in\mathbb{Z}_{N/d}
    \right\},
\end{equation}
where, for $M=N/d$,
\begin{align}
    L_{d,\beta}
    &\triangleq
    \left\langle
        (d,\beta),(0,M)
    \right\rangle_{\mathbb{Z}_N}
    \notag\\
    &=
    \left\{
        \bigl([td]_N,[t\beta+uM]_N\bigr):
        t\in\mathbb{Z}_M,\;
        u\in\mathbb{Z}_d
    \right\}.
    \label{eq:elementary_lagrangian_form}
\end{align}
Then $\mathcal{L}_N=\mathcal{E}_N$.
\end{theorem}

\noindent
The proof is given in Appendix~A.

\subsection{Parameter Interpretation and Special Cases}
\begin{remark}[Roles of $d$, $M$, and $\beta$]
\label{rem:canonical_DD_interpretation}
For each $t\in\mathbb{Z}_M$, the delay slice located at
$\tau=td$ is
\begin{equation}
    \mathcal{G}_t
    \triangleq
    \left\{
        \bigl([td]_N,[t\beta+uM]_N\bigr):
        u\in\mathbb{Z}_d
    \right\}.
    \label{eq:delay_group}
\end{equation}
The parameter $\beta$ controls the slope, or equivalently the shear,
of the grid. The parameter $d$ determines both the number of parallel
DD lines and the spacing between consecutive delay slices, while
$M=N/d$ determines the Doppler spacing within each slice.
\end{remark}

\begin{example}[Zak-OTFS]
\label{ex:zak_support}
Let \(N=N_{\tau}N_{\nu}\). The Zak-OTFS waveform
\cite{mattu2026zakotfs} diagonalizes the aligned rectangular support
obtained with \(d=N_{\tau}\), \(M=N_{\nu}\), and \(\beta=0\), whose
geometry is illustrated in
Fig.~\ref{fig:lagrangian_submodule_geometry}(a):
\begin{align*}
    L_{N_{\tau},0}
    &=
    \left\langle
        (N_{\tau},0),(0,N_{\nu})
    \right\rangle_{\mathbb{Z}_N}
    \\
    &=
    \left\{
        (aN_{\tau},bN_{\nu}):
        a\in\mathbb{Z}_{N_{\nu}},\;
        b\in\mathbb{Z}_{N_{\tau}}
    \right\}.
\end{align*}
\end{example}

\begin{example}[DAFT-OFDM]
\label{ex:daft_ofdm_support}
For an integer \(\alpha\) satisfying \(\gcd(\alpha,N)=1\), the
CP-compatible DAFT-OFDM waveform\footnote{DAFT-OFDM generally uses a
chirp-periodic prefix, which reduces to an ordinary CP for the integer
chirp parameter considered here \cite{erseghe2005aft}.} diagonalizes
the cyclic DD line obtained with \(d=1\), \(M=N\), and
\(\beta=[2\alpha]_N\), whose geometry is illustrated in
Fig.~\ref{fig:lagrangian_submodule_geometry}(c):
\begin{align*}
    L_{1,[2\alpha]_N}
    &=
    \left\langle
        (1,[2\alpha]_N)
    \right\rangle_{\mathbb{Z}_N}
    \\
    &=
    \left\{
        (r,[2\alpha r]_N):
        r\in\mathbb{Z}_N
    \right\}.
\end{align*}
\end{example}
\begin{example}[Sheared support]
For \(N=24\) and \((d,M,\beta)=(4,6,1)\),
\[
    L_{4,1}
    =
    \left\langle
        (4,1),(0,6)
    \right\rangle_{\mathbb{Z}_{24}}
\]
is the sheared grid shown in
Fig.~\ref{fig:lagrangian_submodule_geometry}(b).
\end{example}

\subsection{Counting Lagrangian Submodules}
The representation in  Theorem~\ref{thm:elementary_representation}  also determines the exact number of
Lagrangian submodules in $\mathbb{Z}_N^2$.

\begin{corollary}
\label{cor:lagrangian_enumeration}
Every positive integer $N>1$ admits a unique prime factorization,
which we write as
\(N=\prod_{i=1}^{r} p_i^{a_i}\),
where $p_1,\ldots,p_r$ are distinct primes and
$a_1,\ldots,a_r$ are positive integers. Then the number of Lagrangian
submodules of $\mathbb{Z}_N^2$ is
\begin{equation}
    \left|\mathcal{L}_N\right|
    =
    \prod_{i=1}^{r}
    \frac{p_i^{a_i+1}-1}{p_i-1}.
    \label{eq:lagrangian_enumeration}
\end{equation}
\end{corollary}

\noindent
The proof is given in Appendix~B.

By now, we have identified all maximal diagonalizable DD supports. In the next section, we design the corresponding transmit
waveforms.

\section{Design of Diagonalizing Waveforms}
\label{sec:diagonalizing_waveforms}
\subsection{Closed-Form Waveforms}

\begin{theorem}[Diagonalizing waveforms]
\label{thm:diagonalizing_waveform_basis}
For each $\rho\in\mathbb{Z}_d$ and $m\in\mathbb{Z}_M$, let
$\gamma_{\rho,m}\in\mathbb{C}$ satisfy
$|\gamma_{\rho,m}|=1$. Write $n=dj+\rho$ with
$j\in\mathbb{Z}_M$ whenever $n\equiv\rho\pmod d$, and define
$\boldsymbol{\psi}_{\rho,m}\in\mathbb{C}^N$ by
\begin{equation}
    \psi_{\rho,m}[n]
    =
    \begin{cases}
        \displaystyle
        \gamma_{\rho,m}M^{-1/2}
        e^{\frac{\pi\mathrm{i}}{M}
        [\beta j(j-M)-2mj]},
        & n=dj+\rho,
        \\[2mm]
        0,
        & \text{otherwise}.
    \end{cases}
    \label{eq:explicit_diagonalizing_waveform}
\end{equation}
Collect the above vectors as the columns of
$\mathbf{B}_{d,\beta}\in\mathbb{C}^{N\times N}$ according to
\begin{equation}
    \bigl[\mathbf{B}_{d,\beta}\bigr]_{:,\,md+\rho}
    =
    \boldsymbol{\psi}_{\rho,m},
    \qquad
    \rho\in\mathbb{Z}_d,\;
    m\in\mathbb{Z}_M.
    \label{eq:diagonalizing_waveform_matrix_definition}
\end{equation}
Then $\mathbf{B}_{d,\beta}$ is unitary and
diagonalizes every channel whose DD support is
$L_{d,\beta}$, which is unique up to column permutation.
\end{theorem}

\noindent
The proof is given in Appendix~C.

\subsection{Matrix Implementation and Special Cases}
Let
\(\mathbf{F}_M
\triangleq
M^{-1/2}
[e^{-2\pi\mathrm{i}jm/M}]_{j,m\in\mathbb{Z}_M}\)
denote the unitary \(M\)-point DFT matrix.
Next, define the diagonal quadratic-phase matrix
\(\mathbf{D}_{\beta,M}\in\mathbb{C}^{M\times M}\) and the diagonal phase matrix
\(\mathbf{\Gamma}_{d,M}\in\mathbb{C}^{N\times N}\) through
\begin{align}
    \bigl[\mathbf{D}_{\beta,M}\bigr]_{j,j}
    &\triangleq
    e^{\frac{\pi\mathrm{i}\beta}{M}j(j-M)},
    \qquad
    j\in\mathbb{Z}_M,
    \label{eq:quadratic_phase_matrix}
    \\
    \bigl[\mathbf{\Gamma}_{d,M}\bigr]_{md+\rho,\,md+\rho}
    &\triangleq
    \gamma_{\rho,m},
    \qquad
    \rho\in\mathbb{Z}_d,\;
    m\in\mathbb{Z}_M.
    \label{eq:column_phase_matrix}
\end{align}
Since \(\lvert\gamma_{\rho,m}\rvert=1\), both
\(\mathbf{D}_{\beta,M}\) and \(\mathbf{\Gamma}_{d,M}\) are unitary.
Then we may rewrite $\mathbf{B}_{d,\beta}$ as
\begin{equation}
    \mathbf{B}_{d,\beta}
    =
    \left[
        \left(
            \mathbf{D}_{\beta,M}\mathbf{F}_M
        \right)
        \otimes
        \mathbf{I}_d
    \right]
    \mathbf{\Gamma}_{d,M}.
    \label{eq:waveform_basis_matrix}
\end{equation}

Recall that the transmitted vector is
$\mathbf{x}=\mathbf{B}_{d,\beta}\mathbf{s}$. Arrange the phase-rotated
symbols columnwise as
$\widetilde{\mathbf{s}}
\triangleq\mathbf{\Gamma}_{d,M}\mathbf{s}
=\operatorname{vec}(\widetilde{\mathbf{S}})$, where
$\widetilde{\mathbf{S}}\in\mathbb{C}^{d\times M}$. Applying
$(\mathbf{A}\otimes\mathbf{B})\operatorname{vec}(\mathbf{S})
=
\operatorname{vec}(\mathbf{B}\mathbf{S}\mathbf{A}^{T})$
to \eqref{eq:waveform_basis_matrix} gives
\begin{equation}
    \begin{aligned}
        \mathbf{x}
        &=
        \left[
            \left(
                \mathbf{D}_{\beta,M}\mathbf{F}_M
            \right)
            \otimes\mathbf{I}_d
        \right]
        \operatorname{vec}(\widetilde{\mathbf{S}})
        \\
        &=
        \operatorname{vec}
        \left(
            \widetilde{\mathbf{S}}
            \left(
                \mathbf{D}_{\beta,M}\mathbf{F}_M
            \right)^{T}
        \right)
        =
        \operatorname{vec}
        \left(
            \widetilde{\mathbf{S}}
            \mathbf{F}_M\mathbf{D}_{\beta,M}
        \right).
    \end{aligned}
    \label{eq:parallel_DFT_implementation}
\end{equation}
The matrix product inside $\operatorname{vec}(\cdot)$ shows that the
symbols first undergo the phase rotations in
$\mathbf{\Gamma}_{d,M}$, after which each row of
$\widetilde{\mathbf{S}}$ undergoes an $M$-point DFT followed by
quadratic-phase rotations. Hence, the modulation can be implemented
using $d$ parallel $M$-point FFTs and pointwise phase rotations. Ignoring sample reordering, the resulting complexity is
\begin{equation}
    \mathcal{C}_{d,\beta,\Gamma}
    =
    \mathcal{O}\!\left(
        N\log M
        +
        N\mathbb{I}\{\beta\neq0\}
        +
        N\mathbb{I}\{
            \mathbf{\Gamma}_{d,M}\neq\mathbf{I}_N
        \}
    \right).
    \label{eq:implementation_complexity}
\end{equation}
Here, \(\mathbb{I}\{\mathsf{P}\}\) equals one when \(\mathsf{P}\)
is true and zero otherwise.
Thus, the implementation complexity is determined by the FFT length
$M$ and whether the two diagonal phase matrices are nontrivial.

\begin{example}
\label{ex:zak_waveform}
For \(d=N_{\tau}\), \(M=N_{\nu}\), and \(\beta=0\), choose
\(\mathbf{\Gamma}_{N_{\tau},N_{\nu}}=\mathbf{I}_N\). Since
\(\mathbf{D}_{0,N_{\nu}}=\mathbf{I}_{N_{\nu}}\),
\eqref{eq:waveform_basis_matrix} reduces to
\begin{equation}
    \mathbf{B}_{N_{\tau},0}
    =
    \mathbf{F}_{N_{\nu}}
    \otimes
    \mathbf{I}_{N_{\tau}},
    \label{eq:zak_waveform_basis_matrix}
\end{equation}
which is the Zak-OTFS waveform\cite{mattu2026zakotfs}.
\end{example}

\begin{example}
\label{ex:daft_ofdm_waveform}
For \(d=1\), \(M=N\), and
\(\beta=[2\alpha]_N\), choose
\(\mathbf{\Gamma}_{1,N}=\mathbf{\Gamma}_N\), where
\(\mathbf{\Gamma}_N\) is diagonal pre-chirp matrix. Then
\eqref{eq:waveform_basis_matrix} becomes
\begin{equation}
    \mathbf{B}_{1,[2\alpha]_N}
    =
    \mathbf{D}_{[2\alpha]_N,N}
    \mathbf{F}_N
    \mathbf{\Gamma}_N,
    \label{eq:daft_ofdm_waveform_basis_matrix}
\end{equation}
which is the DAFT-OFDM waveform
\cite{erseghe2005aft}.
\end{example}

\section{Conclusion}
\label{sec:conclusion}
Under CP-based block transmission and assuming that the modulation waveforms form an orthonormal basis, this paper studied exact channel diagonalization for doubly selective channels that does not depend on the path gains. We identified all maximal diagonalizable DD
supports and derived their corresponding waveforms. The resulting waveforms can be implemented using parallel FFTs and phase rotations, with Zak-OTFS and DAFT-OFDM appearing as special cases.
\appendices
\section{Proof of Theorem~\ref{thm:elementary_representation}}
\label{app:proof_elementary_representation}

We prove the two set inclusions separately.

\emph{1) $\mathcal{E}_N\subseteq\mathcal{L}_N$:}
Let $L_{d,\beta}\in\mathcal{E}_N$ and set $M=N/d$.
By definition, $L_{d,\beta}$ is a submodule of
$\mathbb{Z}_N^2$.

Consider two arbitrary elements
\begin{equation*}
    \boldsymbol{\ell}
    =
    (td,t\beta+qM),
    \qquad
    \boldsymbol{\ell}'
    =
    (t'd,t'\beta+q'M)
\end{equation*}
in $L_{d,\beta}$. Their commutation measure is
\begin{align}
    \omega(\boldsymbol{\ell},\boldsymbol{\ell}')
    &=
    \bigl[
        td(t'\beta+q'M)
        -
        t'd(t\beta+qM)
    \bigr]_N
    \notag\\
    &=
    \bigl[
        dM(tq'-t'q)
    \bigr]_N
    =
    0,
\end{align}
because $dM=N$. Thus, all elements of $L_{d,\beta}$ commute
pairwise.

As $t$ ranges from $0$ to $M-1$, the delay coordinate $td$ takes
$M$ distinct values. For each fixed $t$, the Doppler coordinate
$t\beta+qM$ takes $d$ distinct values as $q$ ranges from $0$ to
$d-1$. Therefore, $|L_{d,\beta}|=Md=N$.
Hence, $L_{d,\beta}$ is a Lagrangian submodule, and therefore
$L_{d,\beta}\in\mathcal{L}_N$. This proves
$\mathcal{E}_N\subseteq\mathcal{L}_N$.

\emph{2) $\mathcal{L}_N\subseteq\mathcal{E}_N$:}
Let $L\in\mathcal{L}_N$ and consider the delay projection
\begin{equation*}
    \pi_{\tau}:
    L
    \longrightarrow
    \mathbb{Z}_N,
    \qquad
    \pi_{\tau}(\tau,\nu)
    =
    \tau.
\end{equation*}
Its image is a subgroup of the additive group $\mathbb{Z}_N$. Hence, for a unique
divisor $d\mid N$, $\operatorname{im}(\pi_{\tau})=d\mathbb{Z}_N$.
Set $M=N/d$. Since $d\in\operatorname{im}(\pi_{\tau})$, there exists
an element $(d,\bar{\beta})\in L$.

Let
\begin{equation*}
    \mathcal{K}
    \triangleq
    \ker(\pi_{\tau})
    =
    \left\{
        (0,b)\in L
    \right\}
\end{equation*}
be the set of elements in $L$ with zero delay. By the first
isomorphism theorem, $L/\mathcal{K}\cong d\mathbb{Z}_N$.
Since $|L|=N,\qquad |d\mathbb{Z}_N|=M$, it follows that
\begin{equation}
    |\mathcal{K}|
    =
    \frac{|L|}{|d\mathbb{Z}_N|}
    =
    \frac{N}{M}
    =
    d.
    \label{eq:kernel_cardinality}
\end{equation}

For any $(0,b)\in\mathcal{K}$, the commutation condition with
$(d,\bar{\beta})$ gives
$\omega\bigl((d,\bar{\beta}),(0,b)\bigr)=db\equiv0\pmod N$.
Because $N=dM$, this implies that $b$ is a multiple of $M$.
Therefore,
\begin{equation}
    \mathcal{K}
    \subseteq
    \left\{
        (0,qM):
        0\leq q<d
    \right\}.
    \label{eq:kernel_subset}
\end{equation}
The set on the right-hand side contains exactly $d$ elements.
Together with \eqref{eq:kernel_cardinality}, this gives
\begin{equation}
    \mathcal{K}
    =
    \left\{
        (0,qM):
        0\leq q<d
    \right\}.
    \label{eq:kernel_form}
\end{equation}

Let $\beta=[\bar{\beta}]_M\in\mathbb{Z}_M$.
Then $\bar{\beta}-\beta=q_0M$ in $\mathbb{Z}_N$ for some integer
$q_0$. Since $(0,M)\in\mathcal{K}\subseteq L$, we obtain
$(d,\beta)=(d,\bar{\beta})-q_0(0,M)\in L$.

Now consider any $(a,b)\in L$. Since
$a\in d\mathbb{Z}_N$, there is a unique
$t\in\{0,\ldots,M-1\}$ such that $a=[td]_N$.
It follows that $(a,b)-t(d,\beta)\in\mathcal{K}$.
Using \eqref{eq:kernel_form}, there exists
$q\in\{0,\ldots,d-1\}$ such that
$(a,b)=t(d,\beta)+q(0,M)$.
Therefore,
\begin{equation*}
    L
    \subseteq
    \left\langle
        (d,\beta),(0,M)
    \right\rangle_{\mathbb{Z}_N}
    =
    L_{d,\beta}.
\end{equation*}

By the first part of the proof, $|L_{d,\beta}|=N$.
Since $|L|=N$ and
$L\subseteq L_{d,\beta}$, we conclude that
$L=L_{d,\beta}\in\mathcal{E}_N$. This proves
$\mathcal{L}_N\subseteq\mathcal{E}_N$
and completes the proof.

\section{Proof of Corollary~\ref{cor:lagrangian_enumeration}}
\label{app:proof_lagrangian_enumeration}
If $L_{d,\beta}=L_{d',\beta'}$, equality of their delay projections
implies $d\mathbb{Z}_N=d'\mathbb{Z}_N$ and hence $d=d'$. Set
$M=N/d$. Since both $(d,\beta)$ and $(d,\beta')$ belong to the same
submodule, their difference lies in its zero-delay subgroup
$\{(0,uM):u\in\mathbb{Z}_d\}$. Thus $\beta'-\beta$ is a multiple of
$M$, which means that $\beta'=\beta$ in $\mathbb{Z}_M$. Therefore,
the parametrization in Theorem~\ref{thm:elementary_representation} is
one-to-one. For each divisor $d\mid N$, there are $N/d$ choices of
$\beta$, and the standard divisor-sum formula yields
\begin{equation*}
    |\mathcal{L}_N|
    =\sum_{d\mid N}\frac{N}{d}
    =
    \prod_{i=1}^{r}
    \frac{p_i^{a_i+1}-1}{p_i-1},
\end{equation*}
which is \eqref{eq:lagrangian_enumeration}.

\section{Proof of Theorem~\ref{thm:diagonalizing_waveform_basis}}
\label{app:proof_diagonalizing_waveform_basis}

Because $L_{d,\beta}$ is generated by $(0,M)$ and $(d,\beta)$, it is
sufficient to construct a common eigenbasis of
$\mathbf{W}_N(0,M)$ and $\mathbf{W}_N(d,\beta)$. From
\eqref{eq:weyl_operator},
\begin{equation}
    \bigl[\mathbf{W}_N(0,M)\mathbf{x}\bigr][n]
    =
    e^{2\pi\mathrm{i}n/d}x[n].
    \label{eq:app_WM_sample_action}
\end{equation}
If $\mathbf{x}$ is an eigenvector with eigenvalue $\mu$, then
\begin{equation}
    \left(e^{2\pi\mathrm{i}n/d}-\mu\right)x[n]=0.
    \label{eq:app_WM_componentwise_condition}
\end{equation}
Hence, for some $\rho\in\mathbb{Z}_d$, the vector can be nonzero only
at the indices $n=dj+\rho$, $j\in\mathbb{Z}_M$, with eigenvalue
$\mu_{\rho}=e^{2\pi\mathrm{i}\rho/d}$. Write
$c[j]\triangleq x[dj+\rho]$.

Now require the same vector to satisfy
$\mathbf{W}_N(d,\beta)\mathbf{x}=\lambda\mathbf{x}$. Evaluating the
$n=dj+\rho$ component gives
\begin{equation}
    \lambda c[j]
    =
    e^{2\pi\mathrm{i}\beta\rho/N}
    e^{2\pi\mathrm{i}\beta(j-1)/M}
    c[j-1],
    \qquad j\in\mathbb{Z}_M,
    \label{eq:app_phase_recursion}
\end{equation}
where $j-1$ is interpreted modulo $M$. Repeated application of
\eqref{eq:app_phase_recursion} yields
\begin{equation}
    c[j]
    =
    c[0]\lambda^{-j}
    e^{2\pi\mathrm{i}\beta\rho j/N}
    e^{\pi\mathrm{i}\beta j(j-1)/M}.
    \label{eq:app_phase_recursion_solution}
\end{equation}
The periodicity condition $c[M]=c[0]$ gives
\begin{equation}
    \lambda^M
    =
    e^{2\pi\mathrm{i}\beta\rho/d}
    e^{\pi\mathrm{i}\beta(M-1)}.
    \label{eq:app_eigenvalue_condition}
\end{equation}
Its $M$ solutions may be indexed by $m\in\mathbb{Z}_M$ as
\begin{equation}
    \lambda_{\rho,m}
    =
    e^{2\pi\mathrm{i}\beta\rho/N}
    e^{\pi\mathrm{i}\beta(M-1)/M}
    e^{2\pi\mathrm{i}m/M}.
    \label{eq:app_generator_eigenvalues}
\end{equation}
Substituting \eqref{eq:app_generator_eigenvalues} into
\eqref{eq:app_phase_recursion_solution} gives
\begin{equation}
    c[j]
    =
    c[0]
    e^{\frac{\pi\mathrm{i}}{M}
    [\beta j(j-M)-2mj]}.
    \label{eq:app_closed_form_samples}
\end{equation}
Choosing $c[0]=M^{-1/2}\gamma_{\rho,m}$ produces the waveforms in
\eqref{eq:explicit_diagonalizing_waveform}.

For different values of $\rho$, the waveforms have disjoint supports
and are therefore orthogonal. For a fixed $\rho$, the inner product
of the waveforms indexed by $m$ and $m'$ is
\begin{equation}
    \boldsymbol{\psi}_{\rho,m}^{H}
    \boldsymbol{\psi}_{\rho,m'}
    =
    \frac{1}{M}
    \sum_{j=0}^{M-1}
    e^{2\pi\mathrm{i}(m-m')j/M}
    =
    \delta_{m,m'}.
    \label{eq:app_waveform_orthogonality}
\end{equation}

Since there are $d$ choices of $\rho$ and $M$ choices of $m$, the
construction yields $dM=N$ orthonormal waveforms, which form the
desired waveform $\mathbf{B}_{d,\beta}$ of
$\mathbb{C}^{N}$.
\bibliographystyle{IEEEtran}
\bibliography{references}

\end{document}